\documentclass[twocolumn,aps,prd,amsmath,amssymb]{revtex4}
\bibpunct{}{}{,}{s}{}{,}
\usepackage{graphicx}
\begin{document}
\newcommand{\md}{{\mathrm{d}}}

\title{Effective theory for the cosmological generation of structure}

\author{Martin Bojowald}
\email{bojowald@gravity.psu.edu}
\affiliation{Institute for Gravitation and the Cosmos,
The Pennsylvania State University, 104 Davey Lab, University Park,
PA 16802, USA}

\author{Aureliano Skirzewski}
\affiliation{Centro de F\'isica Fundamental,
Universidad de los Andes, M\'erida 5101, Venezuela}


\begin{abstract}
  The current understanding of structure formation in the early
  universe is mainly built on a magnification of quantum fluctuations
  in an initial vacuum state during an early phase of accelerated
  universe expansion. One usually describes this process by solving
  equations for a quantum state of matter on a given expanding
  background space-time, followed by decoherence arguments for the
  emergence of classical inhomogeneities from the quantum
  fluctuations. Here, we formulate the coupling of quantum matter
  fields to a dynamical gravitational background in an effective
  framework which allows the inclusion of back-reaction effects.  It
  is shown how quantum fluctuations couple to classical
  inhomogeneities and can thus manage to generate cosmic structure in an
  evolving background. Several specific effects follow from a
  qualitative analysis of the back-reaction, including a likely
  reduction of the overall amplitude of power in the cosmic microwave
  background, the occurrence of small non-Gaussianities, and a possible
  suppression of power for odd modes on large scales without parity
  violation.

\bigskip

{\em Keywords}: cosmology, structure formation, effective equations
\end{abstract}

\maketitle

\section{Introduction}

Modern physics aims to explain complex phenomena by tracing them back
to simpler basic principles. This is also true for cosmology, where
the currently quite inhomogeneous universe is thought to have arisen
from a simple, nearly homogeneous initial state very briefly after the
big bang. An important ingredient for realizations of this scenario is
inflation, an early universe phase in which the expansion was
accelerated due to the presence of a postulated inflaton field whose
negative pressure would drive the universe apart. In inflationary
structure formation,\cite{CosmoPert} initial perturbations are
provided by small quantum fluctuations of matter field modes,
including the inflaton field, which are then enlarged during the phase
of accelerated expansion. Out of these initial seeds the current
structure such as galaxies grows by gravitational attraction.

In this way, classical structure may arise from an unstructured vacuum
state.  There must thus be a transition from quantum fluctuations to
classical perturbations, which is one of the fascinating aspects of
this scenario, but also one of the least understood. The success of
inflationary models in comparison with recent observations of
structure in the cosmic microwave background indicates that the
remarkably direct identification $\phi_{\bf k}=\langle\hat{\phi}_{\bf
  k}^2\rangle^{1/2}$ between the amplitude $\phi_{\bf k}$ of classical
perturbations at wave number $\bf k$ and quantum fluctuations of the
inflaton mode $\hat{\phi}_{\bf k}$ describes this process well. It is
an interesting test for the understanding of cosmology as well as
quantum physics to derive this relation, or provide a different
version with the same observational success.

Details of such a relation may well be important observationally. In
fact, the predictions of inflationary cosmology agree with
observations but not in a completely clean way.  While the result of
the general form of a scale-invariant spectrum of inhomogeneities
after inflation is successful, the total amplitude of predicted cosmic
microwave background anisotropies is too high by several orders of
magnitude compared to observations. One may achieve a lower amplitude
by tuning the inflationary model, but this would eliminate a
considerable part of the appeal of inflation. If only some fraction of
the total quantum fluctuations are transferred to classical
inhomogeneities, on the other hand, a reduction of
total power would result.

Also conceptually, it is worthwhile to study the process of structure
formation because there are deep issues related to the measurement
problem of quantum physics. With the usual interpretation of quantum
mechanics several questions immediately come to mind: What causes the
wave function of the inflaton field to collapse, and why is it not the
expectation value $\langle\hat{\phi}_{\bf k}\rangle$ of the
perturbation operator (which would usually be used but would give zero
in an initial vacuum state) but the quantum fluctuation that is
identified with the classical perturbation? Such questions have been
studied by several groups, justifying the outcome by a combination of
different processes to model the quantum-to-classical
transition. First, a matter state evolving in an inflating
background\cite{GuthPi} becomes highly
squeezed.\cite{GSSqueeze,AlbSqueeze} Formally, such a state has
fluctuations close to those of a classical
distribution,\cite{CosmoDecoh,QuantClassNonVac} which can then emerge
after decoherence.\cite{QuantClassCosmo,PrimDecoh} As always,
decoherence is based on the interaction of quantum degrees of freedom
with an environment whose properties are not measured, and thus
presents a coarse-graining process of the total physical system. Not
surprisingly, due to the complexity, explicit decoherence models
require simplifying assumptions in particular in cosmology. A complete
description is lacking; one rather studies a state on a background,
without coupling it to metric inhomogeneities, followed by a
decoherence phase treated separately. While this does show the right
behavior of fluctuations, it is not clear that these are in fact the
precise inhomogeneities coupling dynamically to metric modes.

Such issues indicate that not all crucial physical ingredients of the
situation may have been included yet. One is using the inflaton as a
quantum field in a dynamical universe where its coupling to the
space-time metric and the gravitational field is essential, but not
considered in the process above. In such a context, standard quantum
field theory techniques of fields on dynamical backgrounds or even
quantum gravity to describe the coupling to the metric become very
complicated. Fortunately, though, the set-up of the situation, based
on quantum fluctuations and classical inhomogeneities, indicates that
no strong quantum gravity properties nor technical details of
quantized fields are required for the process. Such situations can
usually be dealt with very powerfully by effective descriptions which,
as we will see in this paper, are in fact quite suitable. 

In this way we will provide a description of properties of the quantum
system through classical equations which are amended by quantum
correction terms or by the inclusion of quantum degrees of freedom.
Such a description is well known from low energy effective actions
used in particle physics to describe perturbative excitations out of
the vacuum of the theory.\cite{PositronEffAc,VacPolEffAc} Related
techniques are used in condensed matter physics in a variety of
different forms. For cosmological purposes, however, this has to be
generalized because the coupling to metric modes is important, and
thus specifying the vacuum state would involve gravitational degrees
of freedom, too, and require quantum gravity.  The techniques we will
use here are general enough to include also metric perturbations in
addition to fluctuations of the inflaton, while reducing to the low
energy effective action in the standard context.\cite{EffAc,Karpacz}
Thus, what we will be using is a proper extension of effective action
schemes to a cosmological context with a dynamical metric.

In this paper our aim is to provide effective equations for a quantum
state of matter on an evolving space-time, including quantum
fluctuations and their back-reaction on the space-time. We focus on a
discussion of the meaning and form of effective equations rather than
the technical derivations or analyses, for which we refer to existing
papers or future work. In a qualitative analysis we will here
highlight several possible effects which show the potential behind
this new type of effective equations for cosmology.

\section{Effective equations}

The basic observation of the required generalization can easily be
illustrated by quantum mechanics of a system with a single degree of
freedom.  Instead of using a representation of states in a Hilbert
space, taking the usual analytical point of view, one can treat
quantum mechanics more algebraically. Akin to algebraic quantum field
theory, one views the algebra of operators and their dynamics as
primary and directly extracts observable information without
specifying states or a quantum representation \footnote{For
  completeness we mention that, alternatively, one can take yet
  another, geometrical viewpoint. It is based on the observation that
  a Hilbert space is an infinite-dimensional phase space with
  coordinates given, e.g., by expansion coefficients $c_j$ of wave
  functions $|\psi\rangle=\sum_jc_j|\psi_j\rangle$ and Poisson
  brackets $\{{\rm Re} c_j,{\rm Im} c_k\}=
  \frac{1}{2\hbar}\delta_{jk}$ which can be derived from the imaginary
  part of the inner product.\cite{GeomQuantMech,ClassQuantMech,Schilling}
  (For simplicity, we
  are ignoring the fact that the quantum phase space arises from the
  projective Hilbert space.) The Schr\"odinger equation for a
  Hamiltonian operator $\hat{H}$ can then be written as Hamiltonian
  equations of motion given by the quantum Hamiltonian $H_Q=\langle
  \hat{H}\rangle$. Indeed, if we choose the basis states $\psi_j$ to
  be given by eigenstates of $\hat{H}$ we have
  $H_Q(c_j)=\sum_jE_j|c_j|^2$, giving equations of motion
  $\frac{\md}{\md t}{\rm Re} c_j = \{{\rm Re} c_j,H_Q\}=
  \frac{E_j}{\hbar}{\rm Im} c_j$ and $\frac{\md}{\md t}{\rm Im} c_j =
  \{{\rm Im} c_j,H_Q\}= -\frac{E_j}{\hbar}{\rm Re} c_j$ which implies
  $\dot{c}_j=-i\hbar^{-1}E_j c_j$ as it follows from the Schr\"odinger
  equation. This phase space formulation brings quantum mechanics
  formally closer to classical mechanics, as far as dynamics is
  concerned, making it sometimes more straightforward to connect
  classical to quantum equations through effective ones. There are,
  however, differences in what are considered observables, an issue we
  will briefly come back to in the end.}.

Dynamical information of a quantum system is contained in the
expectation values $q=\langle\hat{q}\rangle$ and
$p=\langle\hat{p}\rangle$ but, in contrast to a classical system, also
in infinitely many additional variables. The latter represent the
remaining information of a wave function given by all its moments,
which we parameterize in terms of the quantum variables
\begin{equation}
 G^{a,n}:=\langle((\hat{q}-\langle\hat{q}\rangle)^{n-a}
 (\hat{p}-\langle\hat{p}\rangle)^a)_{\rm Weyl}\rangle
\end{equation}
(i.e.\ Weyl ordered operators) for $a=0,\ldots,n$ and integer
$n\geq2$. At $n=2$, for instance, quantum fluctuations
$G^{0,2}=\langle\hat{q}^2\rangle- \langle\hat{q}\rangle^2$ and
$G^{2,2}= \langle\hat{p}^2\rangle- \langle\hat{p}\rangle^2$ as well as
the covariance
$G^{1,2}=\frac{1}{2}\langle\hat{q}\hat{p}+\hat{p}\hat{q}\rangle-
\langle\hat{q}\rangle \langle\hat{p}\rangle$ are among the quantum variables.
This set of variables must be subject to the uncertainty relation
\begin{equation} \label{uncert}
 G^{0,2}G^{2,2}\geq\frac{\hbar^2}{4}+(G^{1,2})^2\,.
\end{equation}

Independently of whether a Heisenberg picture for operators such as
$\hat{q}$ and $\hat{p}$ or a Schr\"odinger picture for wave functions
is used, equations of motion for expectation values take the form
$\dot{q}=\langle[\hat{q},\hat{H}]\rangle/i\hbar$ and
$\dot{p}=\langle[\hat{p},\hat{H}]\rangle/i\hbar$ whose right hand
sides can, for a given Hamiltonian $\hat{H}$, be expressed as a
function of expectation values of $\hat{q}$ and $\hat{p}$ as well as
their quantum variables. For an anharmonic oscillator with Hamiltonian
\[
 \hat{H}=\frac{1}{2m}\hat{p}^2+V(\hat{q})= \frac{1}{2m}\hat{p}^2+
 \frac{1}{2}m\omega^2\hat{q}^2+ \frac{1}{3}\lambda\hat{q}^3
\]
we have, for instance, equations of motion
\begin{eqnarray}
 \frac{{\rm d}}{{\rm d} t}\langle\hat{q}\rangle &=& \frac{1}{m}
\langle\hat{p}\rangle \\
 \frac{{\rm d}}{{\rm d} t}\langle\hat{p}\rangle
&=&-V'(\langle\hat{q}\rangle)-\lambda G^{0,2} \label{eomp}
\end{eqnarray}
correcting the classical force $-V'(q)=-m\omega^2q -\lambda q^2$ by a
fluctuation term which is itself dynamical, i.e.\ changes in time.

\begin{figure}
\begin{center}
\includegraphics[width=7cm]{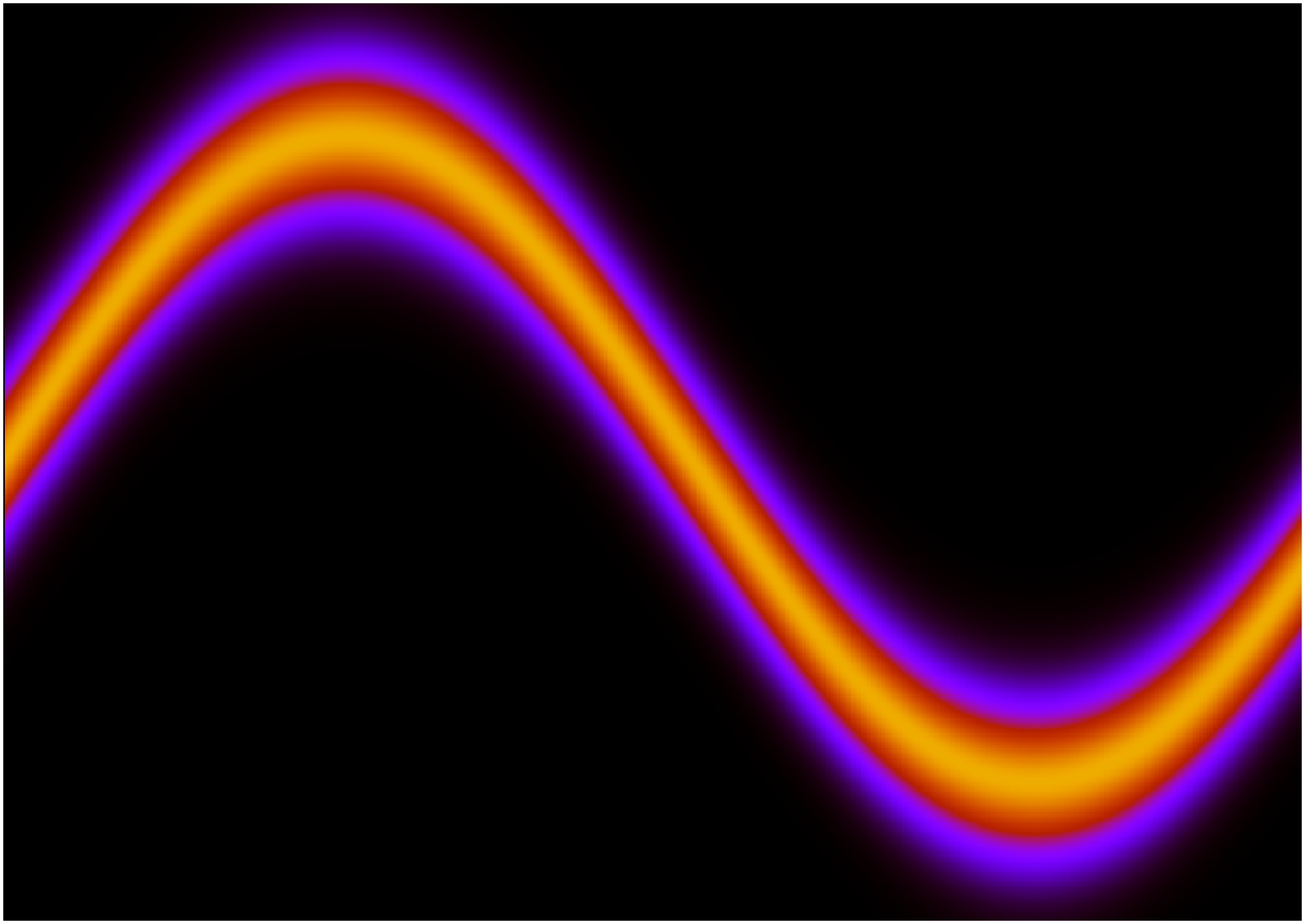}\\
\includegraphics[width=7cm]{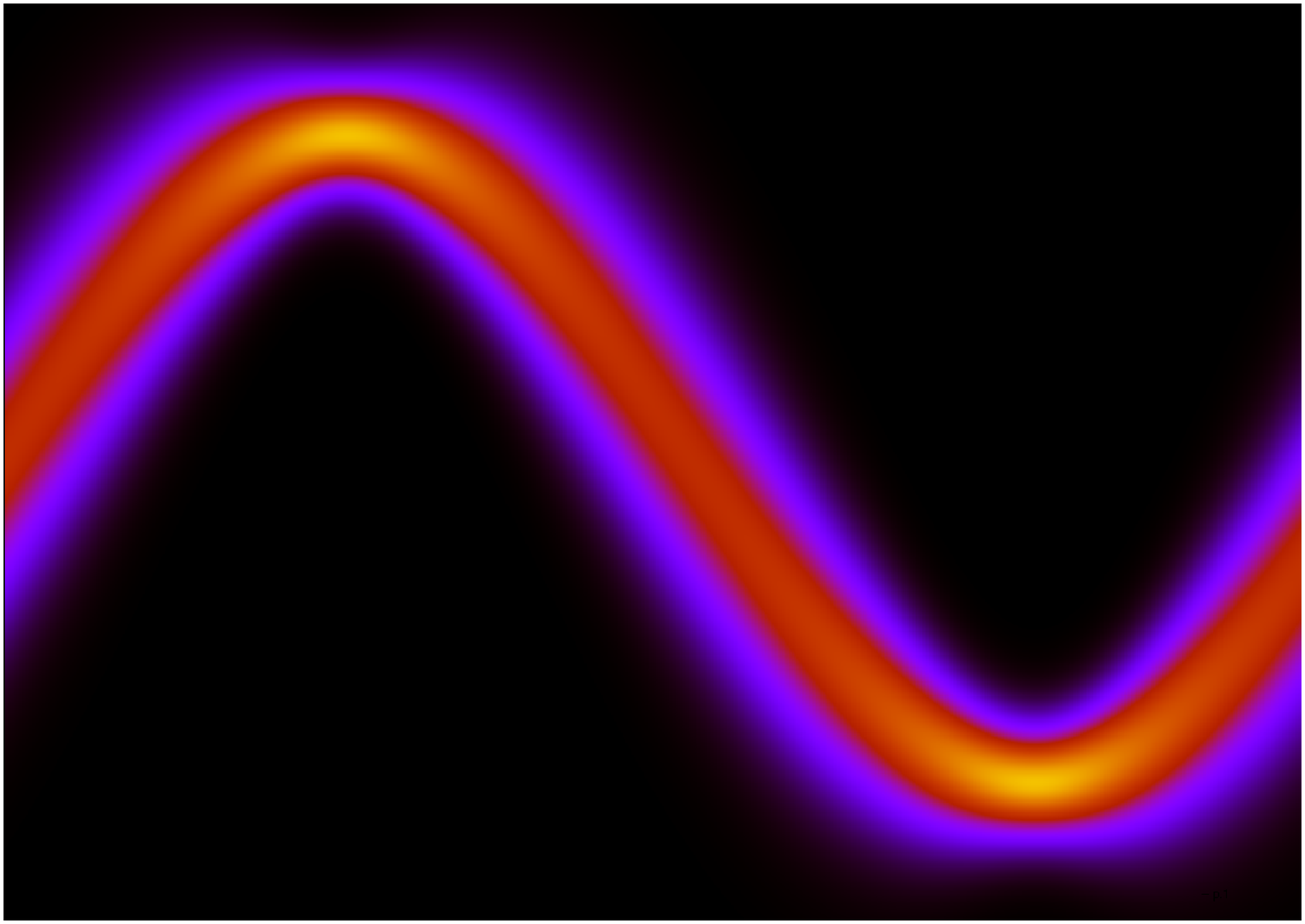}
\end{center}
\caption{An unsqueezed (top) and a squeezed state (bottom) of a
harmonic oscillator, illustrated through the spread $G^{0,2}$, solving
(\ref{Gdot}), around the time dependent expectation value. (The
horizontal axis is time, the vertical one represents $q$.) For the
squeezed state, also fluctuations are time dependent. In a similar
way, fluctuations can be computed in cosmological applications where
squeezing also plays a major role. \label{fig}}
\end{figure}

Equations of motion for quantum variables are thus necessary for a
closed system of equations, but cannot be derived directly from
commutators since they are not expectation values of operators but
also involve products of expectation values. Nevertheless, using the
Leibniz rule one can easily compute equations of motion such as
\begin{equation} \label{Gdot}
 \dot{G}^{0,2} = \frac{\md}{\md t}(\langle\hat{q}^2\rangle-
 \langle\hat{q}\rangle^2) = \frac{\langle[\hat{q}^2,\hat{H}]\rangle}{i\hbar} -
 2q\frac{\langle[\hat{q},\hat{H}]\rangle}{i\hbar}\,.
\end{equation}
Such equations are in some cases easier to solve directly, rather than
computing fluctuations from a complete state. The information gained
is illustrated in Fig.~\ref{fig}.

Moreover, one can express the canonical structure of the full quantum
phase space through Poisson relations derived from commutators, such as
\begin{eqnarray}
 \{G^{0,2},G^{1,2}\} &=& 2G^{0,2}\nonumber\\
 \{G^{0,2},G^{2,2}\} &=& 4G^{1,2}\label{Poisson}\\
 \{G^{1,2},G^{2,2}\} &=& 2G^{2,2}\,.\nonumber
\end{eqnarray}
Similar relations\cite{EffAc} exist between all the $G^{a,n}$.
Alternatively to deriving the equations of motion directly from
expectation values of commutators one can then formulate dynamics
through a quantum Hamiltonian $H_Q=\langle\hat{H}\rangle$. It
determines the usual Hamilton equations of motion not just for
expectation values, $\dot{q}=\{q,H_Q\}$ and $\dot{p}=\{p,H_Q\}$, but
also for quantum variables: $\dot{G}^{a,n}=\{G^{a,n},H_Q\}$.  In
general, i.e.\ unless one is dealing with the quadratic Hamiltonian of
a harmonic oscillator or free particle, $H_Q=\langle\hat{H}\rangle$
will involve quantum variables coupling to the classical ones, which
gives rise to effective corrections to classical equations. A
quadratic Hamiltonian, such as the harmonic oscillator
$\hat{H}=\frac{1}{2m}\hat{p}^2+\frac{1}{2}m\omega^2\hat{q}^2$ does
have quantum variables appearing in its quantum Hamiltonian
\[
 H_Q=\langle\hat{H}\rangle= \frac{1}{2m}(p^2+G^{2,2})+
\frac{1}{2}m\omega^2(q^2+G^{0,2})
\]
but these only provide the zero point energy and do not couple to
expectation values. For an anharmonic oscillator, such as one with the
inclusion of a cubic interaction $\frac{1}{3}\lambda\hat{q}^3$, on the other
hand, we do obtain such coupling terms because
\[
\frac{1}{3}\lambda \langle\hat{q}^3\rangle= \frac{1}{3}\lambda q^3+
\lambda\langle\hat{q}\rangle G^{0,2}+ \frac{1}{3}\lambda G^{0,3}
\]
involves a product of fluctuations and expectation values,
contributing the term $\lambda G^{0,2}$ to (\ref{eomp}), in addition to
the moment $G^{0,3}$ of third order.

Intuitively, these coupling terms describe the motion of the peak of a
wave packet, taking into account the back-reaction of spread and other
deformations on the peak position. In the language of quantum field
theory, one describes the coupled dynamics of $n$-point functions
directly. The situation here is, however, more general since there is
no distinguished state to be used, such as the vacuum state for
$n$-point functions of perturbative low energy quantum field
theory. 

In general, one has to use a suitable class of semiclassical states
for a given regime. They may be difficult to write as explicit states,
but their properties can be seen from analyzing effective equations as
well. Typically not all moments of a semiclassical state are required
at once, and so one can derive the important ones by solving their
equations of motion such as (\ref{Gdot}) along with those of expectation
values. A simple example is again the harmonic oscillator, whose
equations of motion for moments of second order, i.e.\ fluctuations
and the covariance, decouple from the rest. It is straightforward to
derive these equations and see that there is a unique solution with
constant quantum variables saturating the uncertainty relation.

For the harmonic oscillator, coherent states are well known. But the
analysis of equations of motion for moments applies more generally and
provides insights for dynamical coherent states which are difficult to
find as specific wave functions in non-harmonic systems. For instance,
this procedure has recently proven useful in quantum cosmology where
properties of dynamical coherent states describing a non-singular
universe could be determined.\cite{BouncePert,BounceCohStates} This
illustrates limitations to what properties of the very early universe
can be discerned long after the big bang,\cite{BeforeBB,HarmonicCosmo}
and can be extended to more complicated models by a perturbation
analysis.\cite{BouncePot}

\section{Quantum field on a background}

In an analogous way one can describe the dynamical content of a field
theory without using explicit representations of states. Again, each
classical degree of freedom, of which now infinitely many ones exist,
gives rise to an infinite number of quantum degrees of freedom. The
expectation value of the Hamiltonian operator will then play the role
of the Hamiltonian generating evolution of all the $n$-point functions.

We are here primarily interested in the situation of an inflaton field
on an evolving cosmological background.  The Hamiltonian operator of
the inflaton field will thus provide the quantum matter Hamiltonian as
the source in effective cosmological perturbation equations. One can
use the quantum mechanical results described before by performing a
mode decomposition of the field and by quantizing each mode
individually.  (A more rigorous treatment of defining quantum
variables for field theories is possible.\cite{EffectiveEOM}) We
denote the resulting quantum variables as $G^{a,n}_{{\bf
    k}_1,\ldots,{\bf k}_n}$ to indicate the wave vectors ${\bf k}_i$
of each mode operator entering the quantum variable, starting with
$\phi$-modes. Quantum fluctuations then automatically occur in the
generated effective equations, and they do couple to metric modes
since any matter Hamiltonian contains matter as well as geometrical
fields. Through the evolution equations, it is then determined
precisely how these quantum variables couple to classical ones and
lead to classical perturbations in the course of cosmological
expansion.

For a scalar field $\phi$, the inflaton, with momentum $p_{\phi}$ and
potential $V(\phi)=\frac{1}{2}m^2\phi^2$, we have the matter
Hamiltonian
\begin{eqnarray}
 H_{\rm class} &=& \int\md^3x N
\left(\frac{1}{2}q^{-3/2}p_{\phi}^2+ \frac{1}{2} q^{1/2}
\nabla\phi\cdot\nabla\phi+q^{3/2}
V(\phi)\right)\nonumber\\
&& +\int \md^3xp_{\phi} {\bf M}\cdot\nabla\phi\,.
\end{eqnarray}
This Hamiltonian is composed of the kinetic energy
$\frac{1}{2}p_{\phi}^2/q^{3/2}$ with gradient term $\frac{1}{2}
\sqrt{q} \nabla\phi\cdot\nabla\phi$, the potential energy depending on
$V(\phi)$ as well as the momentum flux $p_{\phi}\nabla\phi$. The
function $N$ and the vector $\bf M$ occur in the Hamiltonian because
they determine what is considered the spatial integration over a slice
of space-time in this general relativistic situation. Depending on the
boosting of this slice energy flux occurs in the Hamiltonian if ${\bf
  M}\not=0$.

For the quadratic potential, as typically used in inflation, all terms
in the Hamiltonian are quadratic in the field variables. The theory is
thus free when viewed on a fixed background space-time with metric
components $q$, $N$ and the vector {\bf M} which we used in the
special form
\begin{equation}
  \md s^2 = -N^2\md t^2+ q({\bf x},t)(\md {\bf
    x}+{\bf M}\md t)\cdot (\md {\bf
    x}+{\bf M}\md t)
\end{equation}
suitable for the type of scalar perturbations. (This is relevant for
structure formation in longitudinal gauge; we thus use only spatial
metrics $q_{ij}=q\delta_{ij}$ which are diagonal.)

If the metric is itself dynamical rather than a fixed background,
however, the Hamiltonian is not free. Its terms are no longer
quadratic, and now matter couples to gravity classically but also
quantum mechanically through fluctuations and higher moments. 

To proceed and to show this explicitly, it is useful to write the
functions $\phi$, $p_{\phi}$, $q$, $N$ and all components of ${\bf M}$
in a mode decomposition such as $\phi(t,x)=\bar{\phi}(t)+\sum_{{\bf
    k}\not=0}\phi_{{\bf k}}(t)e^{i{\bf k}\cdot {\bf x}}$. The choice
of background variables, denoted by a bar, depends on the space-time
gauge which, for the metric modes, one usually sets to $\bar{q}=a^2$,
$\bar{N}=a$ (in terms of the scale factor $a$ of a
Friedmann--Robertson--Walker universe) $\bar{{\bf M}}=0$ and
identifies $-q_{{\bf k}}/2a^2=N_{\bf k}/a=:\psi_{\bf k}$ in final
equations. The modes are, however, kept independent to compute
energy-momentum components: energy density modes $\rho_{\bf k}$ are
derived as $a^{-3}\delta H/\delta N_{-{\bf k}}$ and energy flux modes
$V_{i,{\bf k}}$ from $\delta H/\delta M^i_{-{\bf k}}$. Pressure modes
$P_{\bf k}$, which are also important for cosmology, are obtained from
$\delta H/\delta q_{-{\bf k}}$ in a relation which follows from the
definition of pressure as the negative derivative of energy by volume.
%

In a canonical scheme, which is essential for the general theory of
effective systems as described above, irrespective of details of
quantum gravity only the spatial components $q_{ij}$ of the metric
will be quantized while the components $N$ and {\bf M} remain as free
functions (playing the role of Lagrange multipliers of constraints).
The quantum Hamiltonian is thus
\begin{eqnarray*}
 H_Q &=& \int\md^3xN
\left\langle\frac{1}{2}\hat{q}^{-3/2}\hat{p}_{\phi}^2+ \frac{1}{2}
\hat{q}^{1/2}\nabla\hat{\phi}\cdot\nabla\hat{\phi}+\hat{q}^{3/2}
V(\hat{\phi})\right\rangle\\
&& +\int\md^3x{\bf M}\cdot\langle
\hat{p}_{\phi}\nabla\hat{\phi}\rangle
\end{eqnarray*}
clearly showing the non-quadratic nature of the problem.  It is to be
expanded as a series of terms coupling the classical variables to each
other and to quantum variables. Due to the nature of the problem of
interest, i.e.\ relating inflaton fluctuations to classical
inhomogeneities, we consider only quantum fluctuations of matter and
not of the metric. We also ignore correlations between the metric and
matter in this paper.

We have (with $\bar{N}=a$)
\begin{eqnarray*}
 H_Q&=&
H_{\rm class}+ \frac{1}{2a^2}\sum_{\bf k} G^{2,2}_{{\bf k},-{\bf k}}+
\frac{a^2}{2}\sum_{\bf k}(m^2a^2+{\bf k}^2) G^{0,2}_{{\bf k},-{\bf k}}\\&&-
\frac{3}{4a^4} \sum_{{\bf k},{\bf k}'}q_{-{\bf k}-{\bf
    k}'}G^{2,2}_{{\bf k},{\bf k}'}+
\frac{1}{2a^3} \sum_{{\bf k},{\bf k}'}N_{-{\bf k}-{\bf
    k}'}G^{2,2}_{{\bf k},{\bf k}'}\\
&&+\frac{1}{4}\sum_{{\bf k},{\bf k}'}q_{-{\bf k}-{\bf k}'}(3m^2a^2-{\bf k}\cdot
{\bf k}')
G^{0,2}_{{\bf k},{\bf k}'}\\
&&+\frac{1}{2} a\sum_{{\bf k},{\bf k}'}N_{-{\bf k}-{\bf k}'}(m^2a^2-{\bf k}\cdot
{\bf k}')G^{0,2}_{{\bf k},{\bf k}'}\\
&&+i\sum_{{\bf k},{\bf k}'}{\bf k}\cdot {\bf M}_{-{\bf k}-{\bf k}'}
G^{1,2}_{{\bf k},{\bf k}'}
+\cdots
\end{eqnarray*}
where the dots indicate terms of higher order in the perturbations and
those containing quantum correlations between matter and metric
fields. The terms we kept correspond to semiclassical gravity, but
here they are embedded in a larger scheme of the effective quantum
matter and gravity theories. In particular, we will also
derive explicit equations of motion for the fluctuation terms.

Quantum variables are the key new contributions entering effective
equations. In particular, fluctuations
will give non-zero contributions to $\rho_{\bf k}$, $P_{\bf k}$ and
${\bf V}_{\bf k}$ such as
\[
 \frac{1}{a^3}\frac{\delta H_Q}{\delta N_{-{\bf k}}}=\!\frac{1}{2}\!
 \sum_{{\bf k}'}\!\!
\left(\frac{1}{a^6}
 G^{2,2}_{{\bf k}\!-{\bf k}'\!\!,{\bf k}'}\!+\! \frac{m^2a^2\!+\!{\bf k}^{\prime2}\!-{\bf k}\cdot
{\bf k}'}{a^2}
 G^{0,2}_{{\bf k}\!-{\bf k}'\!\!,{\bf k}'}\!\right)\!.
\]
(This is one of the places where quantum field theoretical infinities
can arise. In the effective treatment used here, this can be dealt
with\cite{EffectiveEOM} but will not play a role for the equations
used below.)  Moreover, using the Poisson brackets between quantum
variables, such as
\begin{eqnarray}\{G^{0,2}_{{\bf p}_1,{\bf p}_2},G^{1,2}_{{\bf k}_1,{\bf
      k}_2}\}
&=&\frac{1}{2}(
\delta_{{\bf p}_1,{\bf k}_1}
G^{0,2}_{{\bf p}_2,{\bf k}_2}+\delta_{{\bf p}_1,{\bf k}_2}
G^{0,2}_{{\bf p}_2,{\bf k}_1}
\nonumber\\\nonumber
&&+\delta_{{\bf p}_2,{\bf k}_1}G^{0,2}_{{\bf p}_1,{\bf k}_2}+
\delta_{{\bf p}_2,{\bf k}_2}G^{0,2}_{{\bf p}_1,{\bf k}_1})
\end{eqnarray}
in analogy to (\ref{Poisson}),
we obtain their equations of motion, coupled to
$\psi_{\bf k}$.  

New terms thus result in all equations of motion which are asymptotic
series containing the quantum variables. Including quantum corrections
from fluctuations, we have
\begin{eqnarray}
&&\qquad -{\bf k}^2\psi_{\bf k}-3\frac{\dot{a}}{a}\dot{\psi}_{\bf k}-
3\frac{\dot{a}^2}{a^2}\psi_{\bf k}
= \rho_{\bf k}^{\rm class}\label{const}\\
&&+\frac{1}{2}\sum_{{\bf k}'}(a^{-6}
G^{2,2}_{{\bf k}-{\bf k}',{\bf k}'}+ (m^2-{\bf k}'\cdot({\bf k}-{\bf k}')/a^2)
G^{0,2}_{{\bf k}-{\bf k}',{\bf k}'}) \nonumber
\end{eqnarray}
as the constraint equation whose Lagrange multiplier is $N_{\bf k}$,
\begin{eqnarray}
&&\qquad -\ddot{\psi}_{\bf k}-3\frac{\dot{a}}{a}\dot{\psi}_{\bf k}-
2\left(\frac{\dot{a}}{a}\right)^{.}\psi_{\bf k}-
3\frac{\dot{a}^2}{a^2}\psi_{\bf k} = P_{\bf k}^{\rm class}\label{mot}\\
&&+\frac{1}{2}\sum_{{\bf k}'}(a^{-6}
G^{2,2}_{{\bf k}-{\bf k}',{\bf k}'}-(m^2-{\bf k}'\cdot({\bf k}-{\bf
  k}')/3a^2) G^{0,2}_{{\bf k}-{\bf k}',{\bf k}'})\nonumber
\end{eqnarray}
for the equation of motion of $q_{\bf k}$ and
\begin{eqnarray}
&&i{\bf k}_j(\dot{\psi}_{\bf k}+\frac{\dot{a}}{a}\psi_{\bf k}) = 
{\bf V}_{j,{\bf k}}^{\rm class}
+i\sum_{{\bf k}'}({\bf k}-{\bf k}')_j G^{1,2}_{{\bf k}-
{\bf k}',{\bf k}'}\label{diff}
\end{eqnarray}
for the constraint equation whose Lagrange multiplier is the ${\bf k}$-mode
of the component ${\bf M}_j$. The canonical derivation of these equations
follows that developed\cite{HamPerturb} for a different source of
quantum corrections.

The metric modes are thus explicitly sourced by quantum correlations
between different modes as well as fluctuations in (\ref{const}) and
(\ref{mot}) where $G^{0,2}_{{\bf k}/2,{\bf k}/2}$ and $G^{2,2}_{{\bf
    k}/2,{\bf k}/2}$ contribute. Even if matter inhomogeneities and
thus the density and pressure modes $\rho_{\bf k}^{\rm class}$,
$P_{\bf k}^{\rm class}$ and ${\bf V}_{i,{\bf k}}^{\rm class}$ vanish in an
initial state, quantum fluctuations which must always be present
source and generate perturbations of the classical metric field
$\psi_{\bf k}$.

In this way, quantum fluctuations are the source for
classical perturbations $\psi_{\bf k}$. Moreover, they are themselves
dynamical and change according to the metric perturbations they
generate: we have
\begin{eqnarray}
 \dot{G}_{{\bf p}_1,{\bf p}_2}^{0,2} &=& 2a^{-2}G^{1,2}_{{\bf
     p}_1,{\bf p}_2}\\
 && +4a^{-2}\sum_{{\bf k}}(\psi_{{\bf p}_1-{\bf k}}G_{{\bf p}_2,{\bf k}}^{1,2}+
\psi_{{\bf p}_2-{\bf k}}G_{{\bf p}_1,{\bf k}}^{1,2})\nonumber\\
 \dot{G}_{{\bf p}_1,{\bf p}_2}^{1,2} &=& a^{-2}G^{2,2}_{{\bf p}_1,{\bf
     p}_2}\\
 && +2a^{-2}\sum_{{\bf k}}(\psi_{{\bf p}_1-{\bf k}}G_{{\bf p}_2,{\bf k}}^{2,2}+
 \psi_{{\bf p}_2-{\bf k}}G_{{\bf p}_1,{\bf k}}^{2,2})\nonumber\\
 && -\frac{1}{2}a^2({\bf p}_1^2+{\bf p}_2^2+2m^2a^2)G^{0,2}_{{\bf
     p}_1,{\bf p}_2}\nonumber\\
 && +m^2a^4\sum_{{\bf k}}(\psi_{{\bf p}_1-{\bf k}}G_{{\bf p}_2,{\bf
     k}}^{0,2}+ \psi_{{\bf p}_2-{\bf k}}
 G_{{\bf p}_1,{\bf k}}^{0,2})\nonumber\\
 \dot{G}_{{\bf p}_1,{\bf p}_2}^{2,2} &=&-a^2({\bf p}_1^2+{\bf
   p}_2^2+2m^2a^2)G^{1,2}_{{\bf p}_1,{\bf p}_2}
 \\
 && +2m^2a^4\sum_{{\bf k}}(\psi_{{\bf p}_1-{\bf k}}G_{{\bf p}_2,{\bf
     k}}^{1,2}+\psi_{{\bf p}_2-{\bf k}}
G_{{\bf p}_1,{\bf k}}^{1,2})\nonumber
\end{eqnarray}
for quantum fluctuations of matter on a dynamical space-time.

With infinitely many variables, this is a complicated system of
coupled equations. But one can make several observations already from
the structure of this set. First, through quantum variables, different
metric modes of different wave numbers couple even at the level of
linear metric perturbations with a total quadratic Hamiltonian used
here. In this way, correlations between the modes and thus
non-Gaussianities arise, which are not included in other equations
available so far but will play an increasing role for upcoming
observations.

For a second observation, we now look only at terms containing
$G_{{\bf k},0}$, which are of interest because they derive from
operators linear in the field modes and measure correlations between
the background and inhomogeneities.  They also appear on the right
hand side of metric perturbation equations, e.g.\ $G_{{\bf
    k},0}^{1,2}$ as an addition to the flux in (\ref{diff}) with
similar contributions to $\rho_{\bf k}$ and $P_{\bf k}$. Such a term
appears to correspond to the traditional identification between
inhomogeneities and fluctuations provided that $G_{{\bf
    k},0}^{1,2}=\langle\hat{\phi}_{\bf k}\hat{\bar{p}}_{\phi}\rangle$
can be written as $\bar{p}_{\phi}\langle\hat{\phi}_{\bf
  k}^2\rangle^{1/2}$. If such a relation
would hold, the source term of (\ref{diff}), for instance, could be
written as
\[
{\bf V}_{{\bf k}}^{\rm class}+ i{\bf k}G_{{\bf k},0}^{1,2} = i{\bf
  k}\bar{p}_{\phi} \left(\phi_{\bf k}+ \sqrt{\langle\hat{\phi}_{\bf
    k}^2\rangle}\right)
\]
where the fulctuation could directly take over the role of a classical
perturbation mode $\phi_{\bf k}$.  This form taken exactly, however,
violates the assumption of a Gaussian state (which would have
vanishing correlations) as well as uncertainty relations.  Thus, we
prove that the effective theory of a matter field on an expanding
space-time must result in corrections to the usual direct
identification between matter fluctuations and metric modes.

For other effects, not all the quantum variables are expected to be
equally important, and suitable simplifying truncations are possible.
For instance, we can restrict the set to fluctuations only, i.e.\ only
variables of the form $G_{{\bf k},\pm {\bf k}}$ ignoring correlations.
Then, $\psi_{\bf k}$ together with $G_{{\bf k}/2,\pm{\bf k}/2}^{0,2}$,
$G_{{\bf k}/2,\pm{\bf k}/2}^{1,2}$ and $G_{{\bf k}/2,\pm{\bf
k}/2}^{2,2}$ forms a closed set of equations. This is particularly
relevant because fluctuations are restricted by uncertainty relations
(derived from the pairs $e^{i\theta_1}\hat{p}_{\phi_{\rm k}}\pm
e^{-i\theta_1}\hat{p}_{\phi_{-{\rm k}}}$ and
$e^{i\theta_2}\hat{\phi}_{\rm k}\pm e^{-i\theta_2}\hat{\phi}_{-{\rm
k}}$ of conjugate operators)
\begin{eqnarray*}
 (\pm 2G^{0,2}_{{\bf k},-{\bf k}}+ e^{2i\theta_1}G^{0,2}_{{\bf k},{\bf k}}+
e^{-2i\theta_1}G^{0,2}_{-{\bf k},-{\bf k}})
 (\pm 2G^{2,2}_{{\bf k},-{\bf k}}&&\\
+e^{2i\theta_2}G^{2,2}_{{\bf k},{\bf k}}
+e^{-2i\theta_2}G^{2,2}_{-{\bf k},-{\bf k}}) \geq
\hbar^2\cosh^2(\theta_1-\theta_2)&&\\
+
\left(\pm 2{\rm Re}(e^{i(\theta_1-\theta_2)}G^{1,2}_{{-\bf k},{\bf k}})+
2{\rm Re}(e^{i(\theta_1+\theta_2)}G^{1,2}_{{\bf k},{\bf
    k}})\right)^2
\end{eqnarray*}
for all real $\theta_1$ and $\theta_2$
which follow as in (\ref{uncert}), noting that $\hat{\phi}_{-{\bf
    k}}=\hat{\phi}^{\dagger}_{\bf k}$. Thus, not all $G_{{\bf k},\pm
  {\bf k}}$ can be zero. Nevertheless, $\psi_{\bf k}=0$ with $G_{{\bf
    k},{\bf k}'}=0$ for ${\bf k}\not=-{\bf k}'$ is a consistent
solution, showing that deviations from initial Gaussian states, or
quantum gravitational fluctuations, are required for structure
generation. The vacuum state of a quantum field on a classical
space-time, which would be Gaussian, cannot generate metric
inhomogeneities. 

If this is to be used for inflationary structure formation, there must
be additional source terms in the equations of motion. (Other
scenarios have been proposed.\cite{InflStruc}) The final source not
yet included comes from contributions involving quantum variables of
the metric modes, which would require quantum gravity for their
derivation. While such a calculation would be challenging, it has the
promise of completing the picture of structure formation in the early
universe. In fact, a purely homogeneous space is not consistent with
what one often expects from quantum gravity: a discrete structure of
space-time. Implications of this Planck-scale discreteness also enter
the effective equations through quantum variables involving the metric
operators. They must then appear as additional source terms which
would make exactly homogeneous solutions, where all $\psi_{\bf k}$
vanish, inconsistent. An implementation, which is beyond the scope of
this paper but would follow the same lines, would not only be an
important ingredient for the scenario of structure formation but also
allow tests of specific candidates of quantum gravity by the source
terms of structure they provide.


\section{Discussion}

Effective equations present a complete framework to study the
generation of structure from quantum fluctuations in a way which can
avoid further conceptually involved input from the quantum mechanical
measurement problem. A wave function never occurs explicitly in the
equations, but its physical properties are included through observable
quantities such as its expectation values, fluctuations or
correlations. These variables are more directly related to the
classical ones at least in their mathematical form.

Instead of conceptual problems, we are faced with a computational
problem of analyzing coupled differential equations. Since a priori
infinitely many variables are present, due to the field theoretical
nature but also due to the number of quantum variables describing a
wave function, special solution techniques are required. As in other
examples of effective equations, this usually involves the truncation
of the equations to finitely many ones, based on assumptions for the
magnitude of quantum variables such as the size of fluctuations
compared to that of higher moments. If such a truncation has been
performed, the scheme of solving coupled ordinary differential
equations for expectation values and moments has strong numerical
advantages over solving a partial differential equation for a state
first and then integrating a possibly highly oscillatory semiclassical
state to obtain moments. But also analytically this scheme is highly
economical since it can show several intuitive properties as one is
used to from effective equations.  We have presented two examples for
such considerations of only a truncated subset of quantum variables,
although we did not support this here by a precise estimate of the
ignored terms.

While the system of coupled equations is large and its analysis still
incomplete, several qualitative effects are visible: (i) Correlations
build up during evolution in an indirect process starting from quantum
fluctuations, and then feed classical inhomogeneities.  This results
in a smaller amplitude of inhomogeneities compared to the traditional
identification and could explain the observed discrepancy for the
total power of inhomogeneities.  (ii) States, described by quantum
variables, evolve in complicated ways with all variables coupled to
each other. For instance, correlations between different modes will
arise, implying non-Gaussianity, also at a small level, even from
linear metric perturbations. (iii) Although all quantum variables
contribute as sources of $\psi_{\bf k}$, there is only one fluctuation
term $G_{{\bf k}/2,\pm{\bf k}/2}$ in its equation of motion.  For this
term to exist, the wave number ${\bf k}/2$ must occur in the wave
vector lattice for a given spatial topology. This may not be the case
if we have a compact space for which the ${\bf k}$-space is a
lattice. For a toroidal space, for instance, only $\psi_{\bf k}$ for
even ${\bf k}$ are generated directly while odd modes are suppressed,
most strongly so for small ${\bf k}$ for which it is less likely to
find an existing wave number close to ${\bf k}/2$. This provides a
possible mechanism for the observed suppression of odd modes on large
scales.\cite{WMAPOdd}  Since any of the components of ${\bf k}$ must
be even for ${\bf k}/2$ to be guaranteed to lie on the wave vector
lattice, the relevant parity is mirror rather than point symmetry for
which indeed odd modes are suppressed. With a lattice depending on
spatial topology, a precise comparison with data can reveal
topological properties of space. The mechanism does not require parity
violation\cite{ParityAnomalies} but is, in fact, a
consequence of a parity invariant matter Hamiltonian combined with the
fact that its main terms are quadratic in the matter field.

Regarding interpretational issues, we do not have a sharp transition
from quantum to classical behavior; instead, fluctuations always
remain coupled to classical variables. In some regimes they can be
ignored to an excellent approximation, in which case we obtain a
classical description (akin to ``decoherence without
decoherence''\cite{CosmoDecoh}). This must happen in cosmology:
initially, quantum fluctuations are the only inhomogeneities and seed
classical metric modes. After some time, metric modes grow larger and
fluctuations become less and less relevant in comparison. Their
increasing irrelevance for further evolution is perceived as a quantum
to classical transition. This qualitative behavior still is to be seen
precisely from a detailed analysis, which can shed light on the role
of quantum mechanics in the problem at hand.

At a technical level, those subdominant quantum
variables can then be ignored for an analysis of the equations of
motion. Mathematically this provides an approximation scheme, but from
the physical perspective eliminating some of the quantum variables
from further considerations implies the occurrence of mixed states due
to coarse-graining the degrees of freedom. In fact, not every set of
quantum variables corresponds to a pure state; setting some of them to
zero means that the corresponding state they determine can become
mixed.

The system of differential equations is deterministic and thus appears
to contradict basic quantum mechanics. However, it contains infinitely
many quantum variables and thus requires infinitely many initial
conditions to fully specify a solution. Moreover, unlike the
classical case, not every function on the quantum phase space is an
observable. Not all information about a solution, and thus a quantum
state, is then observationally accessible.  Therefore, unless one
finds theoretical arguments for a unique initial state one can only
describe ensembles of histories which are ultimately compared with
observations in our own universe. In this way, the usual probabilistic
quantum behavior arises and probability distributions appear to
describe our fundamental ignorance.

There are certainly several open issues to be explored within this
scheme, in addition to making the above predictions quantitative. For
instance, a general study of initial states, parameterized by initial
values for the quantum variables, will reveal how sensitive
inflationary results are to that choice. Other questions are also
lurking in the background. In fact, we have treated the metric as a
flat background on which linear perturbations evolve. But in a
complete theory, the whole metric components should be quantized which
can give additional quantum gravity corrections from metric
fluctuations. Some related terms in effective equations, taking into
account quantum gravity effects but not quantum variables, have been
obtained.\cite{InhomEvolve} All this provides the basis for a
systematic investigation of quantum effects of any kind in a combined
theory of gravity and matter.

{\bf Acknowledgements:} We are grateful to Stephon Alexander, Cliff
Burgess, Kate Land and Jo\~ao Magueijo for discussions. MB was
supported in part by NSF grants PHY0554771 and PHY0653127, and by the
Perimeter Institute for Theoretical Physics.


\end{document}